\documentclass[superscriptaddress,twocolumn,prb]{revtex4}
\usepackage{amsmath,graphicx}

\begin{document} 

\title {Evaluating charge noise acting on semiconductor quantum dots in the circuit quantum electrodynamics architecture}
\author{J.~Basset} 
\affiliation{Solid State Physics Laboratory, ETH Zurich, CH-8093 Zurich, Switzerland }
\author{A.~Stockklauser}
\affiliation{Solid State Physics Laboratory, ETH Zurich, CH-8093 Zurich, Switzerland }  
\author{D.-D.~Jarausch} 
\affiliation{Solid State Physics Laboratory, ETH Zurich, CH-8093 Zurich, Switzerland } 
\author{T.~Frey}
\affiliation{Solid State Physics Laboratory, ETH Zurich, CH-8093 Zurich, Switzerland }
\author{C.~Reichl} 
\affiliation{Solid State Physics Laboratory, ETH Zurich, CH-8093 Zurich, Switzerland }
\author{W.~Wegscheider} 
\affiliation{Solid State Physics Laboratory, ETH Zurich, CH-8093 Zurich, Switzerland } 
\author{A.~Wallraff} 
\affiliation{Solid State Physics Laboratory, ETH Zurich, CH-8093 Zurich, Switzerland }
\author{K.~Ensslin} 
\affiliation{Solid State Physics Laboratory, ETH Zurich, CH-8093 Zurich, Switzerland }  
\author{T.~Ihn} 
\affiliation{Solid State Physics Laboratory, ETH Zurich, CH-8093 Zurich, Switzerland }


\begin{abstract}

We evaluate the charge noise acting on a GaAs/GaAlAs based semiconductor double quantum dot dipole-coupled to the voltage oscillations of a superconducting transmission line resonator. The in-phase ($I$) and the quadrature ($Q$) components of the microwave tone transmitted through the resonator are sensitive to charging events in the surrounding environment of the double dot with an optimum sensitivity of $8.5\times10^{-5}~\mbox{e}/\sqrt{\mbox{Hz}}$. A low frequency $1/f$ type noise spectrum combined with a white noise level of $6.6\times10^{-6}$~$\mbox{e}^2/\mbox{Hz}$ above $1$~Hz is extracted, consistent with previous results obtained with quantum point contact charge detectors on similar heterostructures. The slope of the $1/f$ noise allows to extract a lower bound for the double-dot charge qubit dephasing rate which we compare to the one extracted from a Jaynes-Cummings Hamiltonian approach. The two rates are found to be similar emphasizing that charge noise is the main source of dephasing in our system.

\end{abstract}

\maketitle  
Combining semiconductor nanostructures with microwave frequency resonators should allow to realize cavity quantum electrodynamics experiments on a chip~\cite{Raimond01,Wallraff04}. The new possibility being that the qubit may now consist of a single electron \cite{Childress04,Bergenfeldt12,Lambert13} or two-electron spins confined in quantum dots \cite{Petta05,Burkard06,Trif08,Cottet10,Hu12,Jin12}, potentially overcoming the relaxation and dephasing times of actual superconducting qubits.
Additionally, it attracts a lot of interest in the mesoscopic physics community where, for example, photon-mediated non-local electronic transport between separated quantum dots has been predicted \cite{Bergenfeldt12,Delbecq13} and quantum capacitance measurements on single quantum dots have been realized~\cite{Delbecq11,Frey12b}. 
In each of these experiments combining semiconductor quantum dots and superconducting microwave resonators, one element was used to study the properties of the other, i.e.~the resonator allowed to probe the dot's properties \cite{Delbecq11, Frey12, Frey12b, Petersson12, Delbecq13, Toida13, Basset13,Viennot13} or vice versa \cite{Frey11}. 
However, an important milestone remaining to be reached is the strong coupling regime of cavity QED \cite{Raimond01} in which an entangled state between the resonator and the dot is formed. Reaching this regime is challenging with quantum dots due to the lack of control over the decoherence mechanisms limiting charge relaxation and dephasing rates. 
Here, we use the distributed resonator-dot system to quantitatively extract charge fluctuations in the environment surrounding our single-electron GaAs double quantum dot, one source of decoherence in our experiment. We demonstrate a charge sensitivity of the microwave readout at the level of $8.5\times10^{-5}~e/\sqrt{\mbox{Hz}}$ and use this sensitivity to quantitatively probe the low-frequency charge noise of the host heterostructure\cite{Lu03,Reilly07,Cassidy07,Vink07,Buizert08,Muller10,Prance12,Takeda13}. The achieved sensitivity is of the same order as that of typical QPC based charge detectors\cite{Field93,Buks98,Sprinzak00,Gustavsson06,Hanson07,Ihn10,Kung12} and compares favorably to recent lumped element LC resonator techniques which obtained a charge sensitivity of $~6.3\times 10^{-3} e/\sqrt{\mbox{Hz}}$~\cite{Colless13}.

More importantly, the analysis presented here allows us to infer a lower bound for the dephasing rate originating from the low frequency charge noise in the vicinity of the double-dot. The inferred value is similar to that extracted from an analysis of frequency shifts and linewidth broadenings based on a master equation simulation~\cite{Frey12,Petersson12,Basset13}. This emphasizes that charge noise is the main source of dephasing in our double quantum dot-based charge qubit system.

\begin{figure}[htbp]
  \begin{center}
		\includegraphics[width=8.4cm]{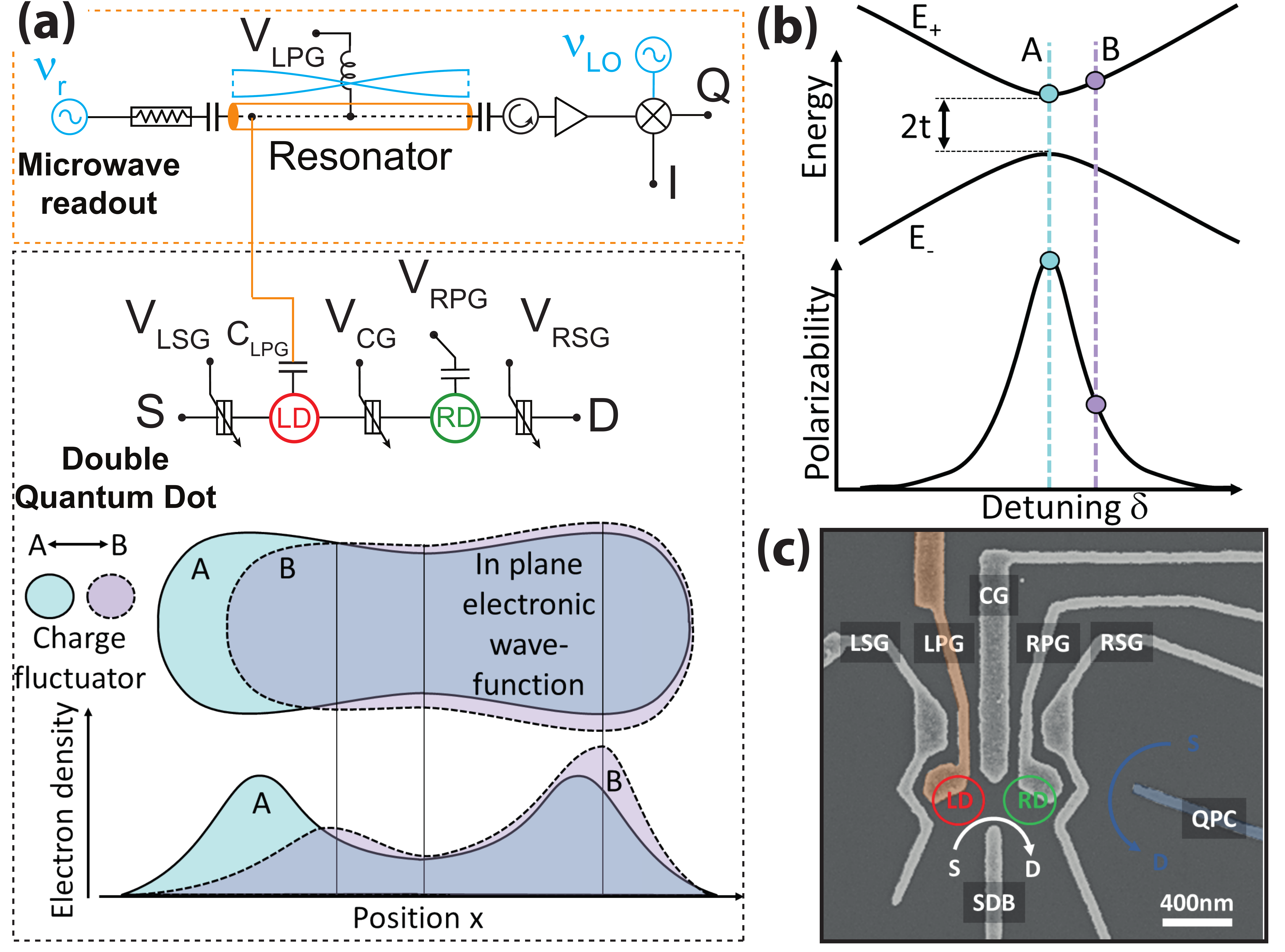}
	\end{center}
  \caption{(a) Circuit diagram representation of the double quantum dot (bottom panel) coupled to the resonator (top panel). The double quantum dot is tuned with gate voltages $V_{\mathrm{LPG}}$, $V_{\mathrm{RPG}}$, $V_{\mathrm{CG}}$, $V_{\mathrm{SDB}}$, $V_{\mathrm{LSG}}$, $V_{\mathrm{RSG}}$. It is connected to the resonator via the capacitor $C_{\mathrm{LPG}}$. The resonator is driven with a microwave signal at frequency $\nu_{\mathrm{r}}$. The transmitted signal passes through a circulator, is amplified and mixed with the local oscillator $\nu_{\mathrm{LO}}$ to obtain the field quadratures I and Q. A charge fluctuator in the vicinity of the double dot affects its wave function and thereby its polarizability. (b) The charge fluctuator as represented in (a) changes the dot detuning $\delta$ (top graph) and accordingly its polarizability (bottom graph). (c) Scanning electron microscope picture of a double quantum dot gate design similar to the one used in the experiment. The gate extending from the resonator is colored in orange. The gate QPC used for reading out the charge state of the double dot is colored in blue.}
  \label{Fig1}
\end{figure}

The sample consists of a two-dimensional electron gas (2DEG) formed $90$ nm below the surface of a GaAs/GaAlAs heterostructure. A split-gate device is defined by electron-beam lithography enabling the formation of a double quantum dot when suitable negative gate voltages are applied (see Fig.~\ref{Fig1}c and bottom panel of Fig.~\ref{Fig1}a). Next to it, a $200$ nm thick superconducting coplanar waveguide resonator made of aluminum is patterned on a region where the 2DEG has been etched away to avoid losses (see top panel of Fig.~\ref{Fig1}a and Refs.~\onlinecite{Frey12,Frey12b,Toida13,Basset13}). The coupling of the resonator to the external microwave feed lines is realized through finger capacitors realizing an over-coupled resonator \cite{Goeppl08}. The resonance frequency and loaded quality factor obtained with this geometry are $\nu_0=6.76$ GHz and $Q_L \approx 920$, respectively. The coupling between the double quantum dot and the resonator excitations is mediated by the left plunger gate (LPG) extending from the resonator to the left quantum dot (orange colored gate in Figs.~\ref{Fig1}a and c) \cite{Basset13}. A quantum point contact is fabricated on the right hand side of the double dot (blue colored gate QPC in Fig.~\ref{Fig1}c). The sample is then operated in a pulse-tube based dilution refrigerator having a base temperature of approximately $10$ mK.

We form a double quantum dot potential by suitably energizing all gates presented in Fig.~\ref{Fig1}c. For its characterization, we record the current flowing through the double quantum dot from source (S) to drain (D) to recover a typical hexagon-shaped charge stability diagram in the many-electron regime~\cite{VanderWiel02,Hanson07,Ihnbook}. We use the nearby quantum point contact (QPC colored in blue in Fig.~\ref{Fig1}c) as a charge sensor \cite{Field93,Elzerman03} to identify the $(0,1)\leftrightarrow(1,0)$ [(N,M) notation corresponds to  N (M) electrons confined in the left (right) dot] transition of interest for which only one electron resides in the DQD and no dc current is measurable~\cite{Field93,Elzerman03,Basset13}.

We then probe the quantum dot's charge state using the resonator by applying a coherent microwave tone at frequency $\nu_0$ to the resonator and extract the amplitude $A$ and phase $\phi$ of the transmitted signal from the measured field quadratures $I$ and $Q$, as $Ae^{i\phi}=I+iQ$ in a heterodyne detection scheme \cite{Wallraff04}. 
As already pointed out in Refs.~\onlinecite{Frey12,Petersson12,Basset13}, the observed amplitude and phase variations when sweeping along the detuning axis $\delta$ allow us to extract the tunnel coupling $t$ between the dots and to estimate the dephasing rates $\gamma_{\varphi}/2\pi$ of the double quantum dot system. This analysis uses a master equation simulation based on the Jaynes-Cummings Hamiltonian~\cite{Frey12} and allows us to find dephasing rates in the GHz range (for more information, see Refs.~\onlinecite{Frey12,Basset13}). 

\begin{figure}[htbp]
  \begin{center}
  	\includegraphics[width=8.4cm]{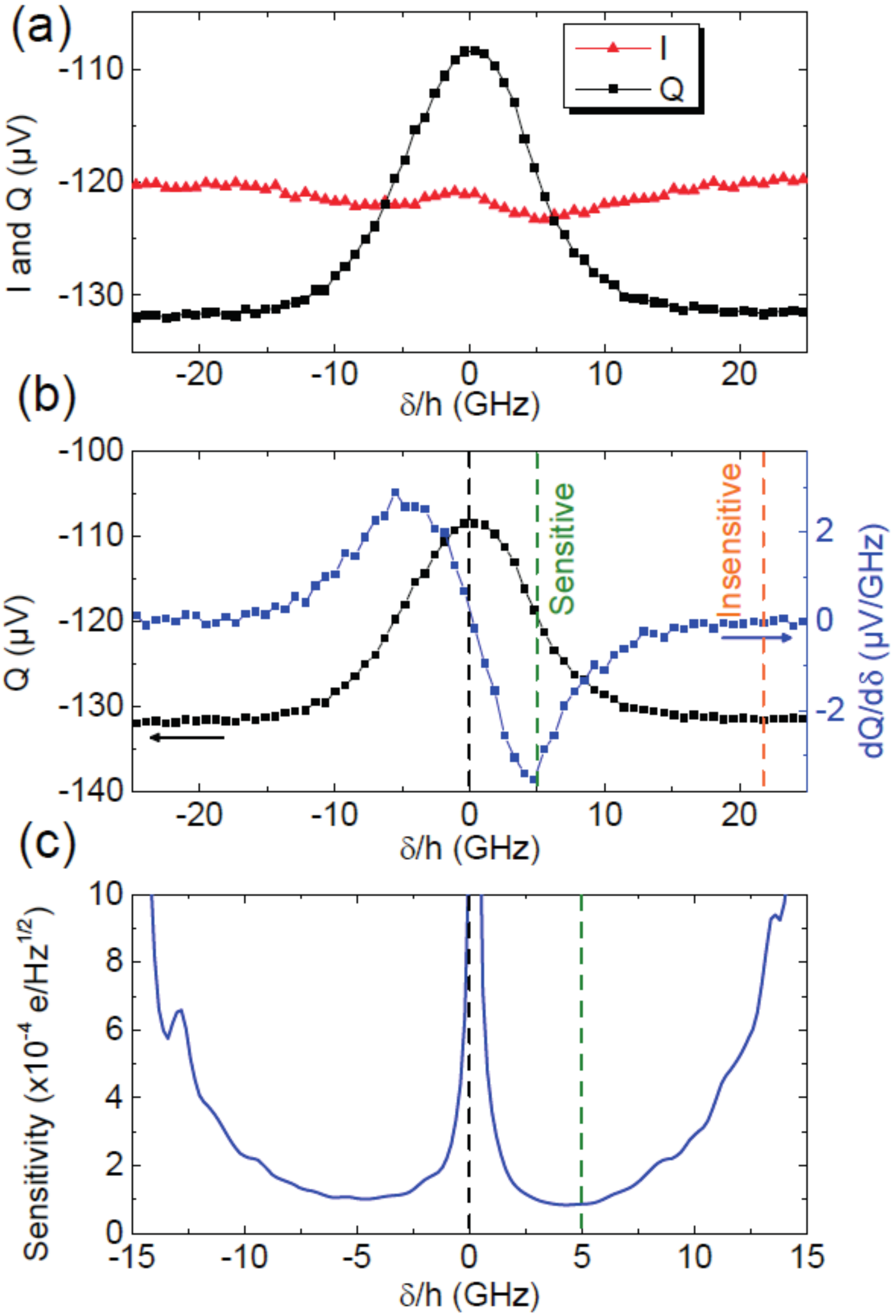}
	\end{center}
  \caption{(a) Measured in-phase $I$ and out-of-phase $Q$ components of the signal transmitted through the resonator vs the detuning $\delta$. (b) Measured quadrature signal $Q$ and  numerical derivative $\partial{Q}/\partial{\delta}$ of $Q$ vs detuning $\delta$. (d) Computed charge sensitivity along $\delta$.}
  \label{Fig2}  
\end{figure}

In the following, we choose to directly measure the quadratures $I$ and $Q$ to extract the charge noise. A typical measurement of $I$ and $Q$ along the detuning line $\delta/h$ is shown in Figs.~\ref{Fig2}a and b for the case $2t/h\nu_0>1$. The conversion of the applied plunger gate voltage to frequency follows from the lever arm consistently extracted from finite bias triangles \cite{VanderWiel02} and electronic temperature broadening of the QPC charge detection linewidth \cite{DiCarlo04}.

Here we use the non-linear dependence of the $Q$ quadrature on the average charge occupancy of the left dot along the detuning line $\delta$ for charge detection. We restrict our analysis to the $Q$ quadrature  because of the small signal to noise ratio of the $I$ component (see Fig.~\ref{Fig2}a).

When tuning across the charge degeneracy line $\delta$, the electron distribution is shifted from the left dot to the right dot. This leads to a change of the energy separation of the qubit states and of its polarizability or quantum admittance~\cite{Petersson10,Frey12b} (see Fig.~\ref{Fig1}b). This scenario can either be realized by tuning the gate voltages $V_{\mathrm{LPG}}$ and $V_{\mathrm{RPG}}$ or alternatively by a fluctuating charged impurity in the vicinity of the double dot (see bottom panel of Fig.~\ref{Fig1}a and the associated changes in energy and polarizability in Fig.~\ref{Fig1}b). We make use of this latter sensitivity to measure the charge noise acting on our double-dot structure.

We define the charge sensitivity of the $Q$ quadrature at a particular detuning $\delta_0$ as
\begin{eqnarray}
\left[\frac{\Delta q}{\sqrt{\Delta \nu}}\right]_{Q} &=& \frac{e}{E_{C}}\frac{\sqrt{S_{Q}}}{\partial{Q}/\partial{\delta}}\Bigg|_{\delta_0}.
\label{ChargeSens}
\end{eqnarray}

In this expression, $S_{Q}$ corresponds to the noise spectral density of the quadrature $Q$ averaged over the measurement bandwidth. The derivative $\partial{Q}/\partial{\delta}$ defines how much the quadrature signal changes with respect to a detuning. $E_c\approx2.4$~meV  is the charging energy of a single dot \cite{Petersson10,Viennot13}. For the sensitivity calculation, we find $S_Q=\sigma_Q^2/\Delta \nu$ with $\Delta \nu =1.53$~Hz the bandwidth of the measurement and $\sigma_Q$ the standard deviation of the measured signal $Q$. In particular, for the setup used in this experiment 
$S_{Q}=5.9\times10^{-23}~\mbox{mV}^2/$~Hz at the input of the amplification chain which corresponds to the noise added by the low temperature HEMT based amplifier. This amplifier has a $[1-14]$~GHz bandwidth, a gain of $39$~dB and a noise temperature of $6$~K. 

In the following we restrict the discussion to the case where $2t/h\nu_0>1$ for simplicity. As a first step we measure $Q(\delta)$ and numerically compute the partial derivative $\partial{Q}/\partial{\delta}$ shown in Fig.~\ref{Fig2}b. 
This allows us to estimate the corresponding sensitivity along the entire detuning axis as defined in Eq.~(\ref{ChargeSens}) (see Fig.~\ref{Fig2}c). The sensitivity shown here has an optimum value of $8.45\times10^{-5}~e/\sqrt{\mbox{Hz}}$. These minima typically arise in the intervals $[-10,-5]$ and $[5,10]$~GHz of the $\delta$ axis~\cite{fn1}. For these detuning ranges, a small change in charge occupation gives rise to a strong change in the quadrature signal. On the other hand, the sensitivit is strongly reduced when the derivative is zero.

\begin{figure}[htbp]
  \begin{center}
		\includegraphics[width=8.5cm]{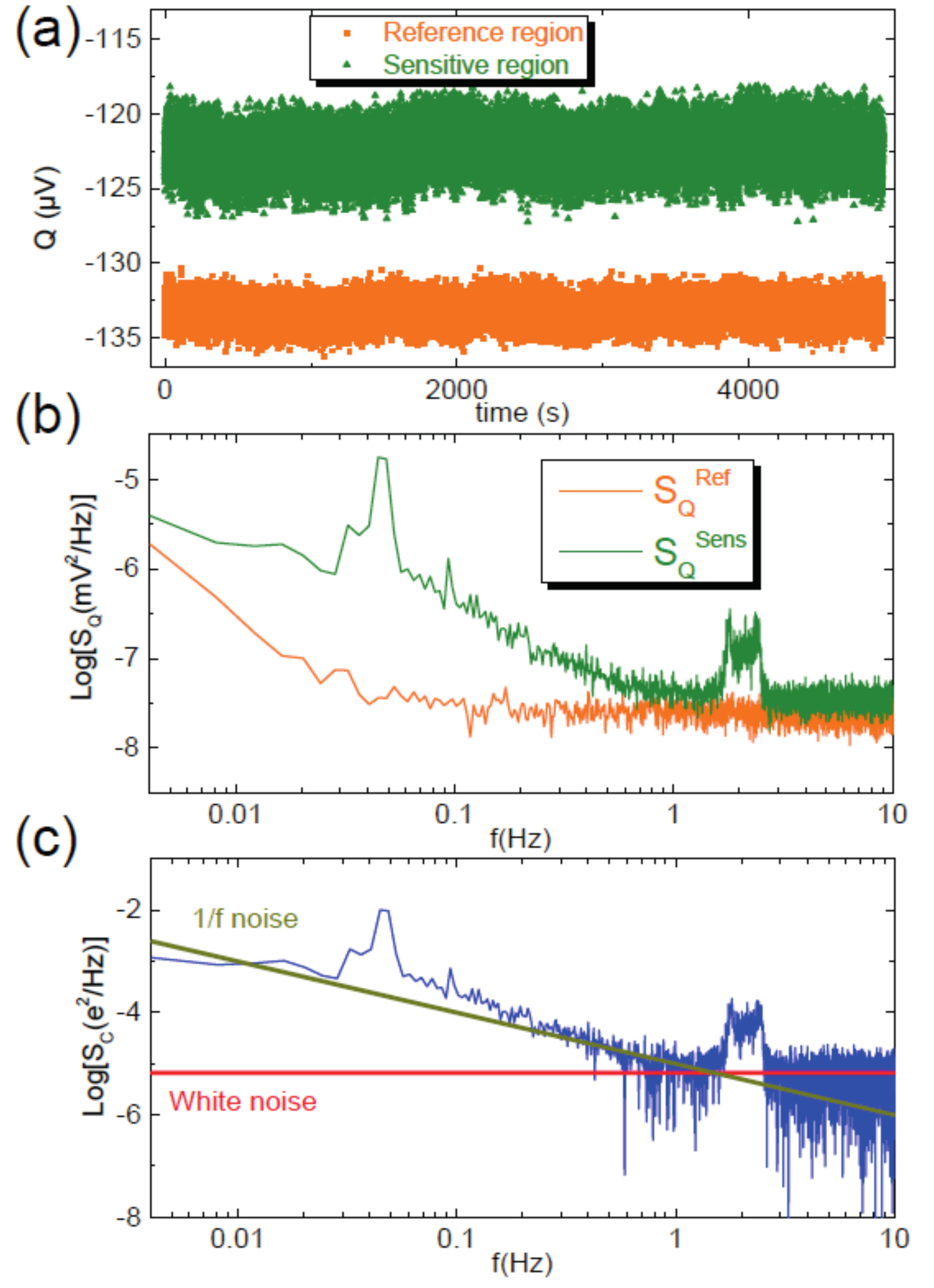}
	\end{center}
  \caption{(a) Quadrature $Q$ time traces for both reference and sensitive gate voltage settings respectively indicated by yellow and green vertical dashed lines in Fig.~\ref{Fig2}. (b) $Q$ noise spectrum calculated from the curves shown in (a). (c) Charge noise spectral density $S_C$ of the surrounding environment of the double dot extracted from $Q$ datasets. The two straight lines are guides to the eye corresponding to a white noise above $1$~Hz or a $1/f$ divergence of the noise at low frequency, respectively.}
  \label{Fig3}  
\end{figure}

We now use this estimate to compute the charge noise spectral density of the surrounding environment into which the double quantum dot is embedded.
To do so, we acquire time traces of $Q$ for two different detuning values \cite{Vink07,Buizert08,Takeda13}. The first detuning value $\delta_1$, highlighted by vertical orange dashed lines in Fig.~\ref{Fig2}, only carries information about the intrinsic noise of the measurement setup. The second value of detuning $\delta_2$ instead, highlighted by vertical green dashed lines in Fig.~\ref{Fig2}, is in addition sensitive to charge fluctuations in the sample. The time traces are acquired for $4920$ seconds with a sampling rate of $17$~Hz (see Fig.\ref{Fig3}a). In order to improve the precision of the extracted noise spectral density we decomposed the time traces into $20$ traces of equal lengths. A discrete Fourier spectrum with coefficients $f_i$ was computed and squared $|f_i|^2=S_i$  for each of these $20$ equivalent datasets and averaged. This leads to the noise spectra shown in Fig.~\ref{Fig3}b. Both spectra exhibit a white noise region above $1$~Hz and a $1/f$ behavior at low frequency. The sensitive region shows a higher noise level and reveals more features than the reference signal.

In order to extract the charge noise spectrum from these measurements we compute
\begin{eqnarray}
S_C(e^2/Hz)&=&\left(\frac{e} {E_C}\right)^2\left(\frac{\partial \delta} {\partial Q}\right)^2\Bigg|_{\delta_2} \left(S_{Q}^{Sens}-S_{Q}^{Ref}\right)
\end{eqnarray}
The result of this calculation is shown in Fig.~\ref{Fig3}c. We notice a sample specific white noise level at high frequency of $6.64 \pm 0.04 \times10^{-6}~e^2/$~Hz and a standard $1/f$ component below $1$~Hz (slope $\sigma_{C}^2=9.91\times 10^{-6}~e^2/Hz$) consistent with previous results obtained with a quantum point contact charge detector on similar heterostructures\cite{Vink07,Buizert08,Takeda13}. Note that unavoidable additional structures occur around $2$~Hz and $0.05$~Hz as a result of either a periodic charge noise in the host material or the interaction of the background charges with the measurement setup fluctuations. Additionally, one can estimate the equivalent detuning noise at $1$~Hz~\cite{Petersson10b,Viennot13}: $S_{\delta}=(E_C/e)^2 S_{C}=57.1~\mu eV^2/$Hz$~=3.3$~GHz$^2/$~Hz.

Using a semiclassical model for dephasing at second order in the charge fluctuation~\cite{Ithier05,Petersson10b,Viennot13}, one can relate the detuning noise $S_{\delta}$ to the double quantum dot charge qubit (energy $\Omega(\delta,t)=\sqrt{\delta^2+(2t)^2}$) dephasing rate via the relation $\gamma_{\varphi}/2\pi\approx\frac{d^2\Omega}{d\delta^2}|_{\delta=0}~S_{\delta}(\nu=1~\rm{Hz})=S_{\delta}(\nu=1~Hz)/2t$.
An estimate of this rate based on the charge noise extracted in this paper leads to $\gamma^C_{\varphi}/2\pi=0.4$~GHz. 

This number is of the same order of magnitude as that extracted from a Jaynes-Cummings analysis of the microwave response, for which we obtained $\gamma_{\varphi}/2\pi\approx0.6$~GHz~\cite{Frey12,Petersson12,Toida13,Basset13} highlighting the relevance of the $1/f$ charge noise to explain the dephasing of quantum dot based charge qubit made with semiconductor heterostructures~\cite{fn2}. Additionally the situation $\gamma^C_{\varphi}/2\pi \lesssim \gamma_{\varphi}/2\pi$ demonstrates that the additional structures revealed in the noise measurements (peaks around $2$~Hz and $0.05$~Hz) and/or the higher frequency spectrum may play a role in the effective dephasing.

At this point, it is interesting to calculate the maximum value of the $1/f$ charge noise slope for which the dephasing rates would become smaller than the coupling to the resonator $\gamma_{\varphi}/2\pi~<~g/2\pi$. For $g/2\pi=25$~MHz as extracted from the experiment, this condition requires $\sigma_{C}^2~<~0.59\times 10^{-6}~e^2/Hz$ which is only an order of magnitude below the values found in our particular heterostructure.

In summary, we have shown that the interaction between a superconducting resonator and a double quantum dot which manifests itself as changes in the quadrature signal transmitted through the resonator can be used as a sensitive probe of charge fluctuations in the host material.

The coupling between the resonator and the double quantum dot gives rise to frequency shifts and linewidth broadenings of the resonator spectra as already pointed out in Refs.~\onlinecite{Frey12,Petersson12}. While these observables strongly depend on the geometry of the sample and on its coherence properties and cannot be changed easily, the quality factor of the resonator instead can be increased, though limited by piezoelectric effects in GaAs. An equivalent frequency shift will lead to a stronger quadrature signal change, the higher the quality factor. By undercoupling the resonator, quality factors of $10^4$ have been demonstrated on a GaAs wafer \cite{Frey11}. This factor of $10$ increase compared to the presently realized resonator would improve the $Q$ sensitivity by a factor of $10$, though limiting the maximum bandwidth of the measurement scheme set by the decay rate of the resonator $\kappa/2\pi=\nu_0/Q_L$~\cite{Colless13}. One would then reach the best sensitivities achieved with state of the art quantum point contact charge detectors \cite{Lu03,Reilly07,Cassidy07,Muller10}. It would be even more interesting to exchange the substrate to avoid limitations due to piezoelectric effects in GaAs. In this regard, Si or Sapphire substrates are natural candidates and have already demonstrated quality factors as high as $10^5-10^6$ at $6$~GHz. In this situation, GaAs quantum dots could be replaced by either InAs nanowires~\cite{Petersson12}, carbon nanotubes~\cite{Delbecq11,Viennot13,Viennot14} or Si/SiGe~\cite{Prance12} quantum dots. Finally, in future experiments, the use of Josephson-based parametric amplifiers~\cite{Castellanos08,Eichler14} instead of HEMT-based amplifiers working in the $GHz$ domain will allow to drastically increase the signal to noise ratio and as a result imcrease the charge sensitivity by several orders of magnitude.

More importantly, we found a slope of the $1/f$ charge noise at the level of $9.9\times10^{-6}$~$\mbox{e}^2/\mbox{Hz}$ below $1$~Hz in our device, typical of GaAs/GaAlAs heterostructures. This noise level translates to a dephasing rate which is slightly smaller than that determined experimentally using the Master equation approach applied to the Jaynes-Cummings model~\cite{Basset13}. This inequality highlights that the background charge fluctuations leading to the $1/f$ charge noise spectrum are the main but not the only ingredients necessary to explain the fast dephasing rates observed in our experiment.

We acknowledge valuable discussions with Takis Kontos, J\'er\'emie Viennot, Jonas Mlynek, Christian Lang, Yves Salath\'e, Lars Steffen, Sarah Hellm\"uller and Fabrizio Nichele. This work was financially supported by the Swiss National Science Foundation through the National Center of Competence in Research "Quantum Science and Technology," and by ETH Zurich.


\begin{thebibliography}{99}
\bibitem{fn1} The exact position in detuning depends on the tunnel coupling between the dots and on the relation $\nu_0-\gamma_{\varphi}/2\pi\leq\delta/h\leq\nu_0+\gamma_{\varphi}/2\pi$, with $\gamma_{\varphi}/2\pi$ the dephasing rate, that has to be satisfied.
\bibitem{fn2} It is not known today which fluctuating background charges, present inside the material or at the surface of the substrate, contribute the more to the observed $1/f$ spectrum.
\end{thebibliography}
\end{document}